\documentclass[aps,prd,preprintnumbers,showpacs,nofootinbib]{revtex4}
\usepackage{graphicx}
\begin{document}
\title{Comparison of high-energy galactic and atmospheric tau neutrino flux}
\author{H. Athar$^{1,2,}$\footnote{E-mail: athar@phys.cts.nthu.edu.tw},
 Kingman Cheung$^{1,}$\footnote{E-mail: cheung@phys.cts.nthu.edu.tw},
Guey-Lin Lin$^{2,}$\footnote{E-mail: glin@cc.nctu.edu.tw}, Jie-Jun
Tseng$^{2,}$\footnote{E-mail: geny.py86g@nctu.edu.tw}}
\affiliation{$^1$Physics Division, National Center for Theoretical Sciences,
Hsinchu 300, Taiwan \\
$^2$Institute of Physics, National Chiao Tung University, Hsinchu
300, Taiwan}
\date{\today}

\begin{abstract}
We compare the tau neutrino flux arising from the galaxy and the
earth atmosphere for $10^{3} \leq E/\mbox{GeV} \leq 10^{11}$. The
intrinsic and oscillated tau neutrino fluxes from both sources are
calculated. The intrinsic galactic $\nu_{\tau}$ flux ($E \geq
10^{3}$ GeV) is calculated by considering the interactions of
high-energy cosmic-rays with the matter present in our galaxy,
whereas the oscillated galactic $\nu_{\tau}$ flux is coming from
the oscillation of the galactic $\nu_{\mu}$ flux. For the
intrinsic atmospheric $\nu_{\tau}$ flux, we  extend the validity
of a previous calculation from $E\leq 10^{6}$ GeV up to $E \leq
10^{11}$ GeV. The oscillated atmospheric $\nu_{\tau}$ flux is, on
the other hand, rather suppressed. We find that, for $10^{3} \leq
E/\mbox{GeV} \leq 5\cdot 10^{7}$, the oscillated $\nu_{\tau}$ flux
along the galactic plane dominates over the maximal intrinsic
atmospheric $\nu_{\tau}$ flux, i.e., the flux along the horizontal
direction. We also briefly mention the presently envisaged
prospects for observing these high-energy tau neutrinos.
\end{abstract}
\preprint{NCTU-HEP-0105} \preprint{NSC-NCTS-011218}
 \maketitle

\section{Introduction}

Searching for high-energy tau neutrinos ($E \geq 10^{3}$ GeV) will
yield quite useful information about the highest energy phenomenon
occurring in the universe \cite{Halzen:2001ty}. The same search
may also provide evidence for physics beyond the standard model
\cite{Fukuda:1998mi}. The latter is suggested by the recent
measurements of the atmospheric muon neutrino deficit, though yet
there is no observation of oscillated atmospheric tau neutrinos at
a significant confidence level \cite{Fukuda:2000np}.
Interestingly, it is also only recently that we have the first
evidence of existence of tau neutrinos \cite{Kodama:2000mp}.

The high-energy tau neutrinos can be produced in $pp$  and
$p\gamma $ interactions taking place in cosmos. These interactions
produce unstable hadrons that decay into tau neutrinos. In this
paper, we mainly concentrate on $pp$ interactions  and will only
briefly comment on $p\gamma $ interactions as source interactions
for producing high-energy  tau neutrinos. There can be several
astrophysical sites where the $pp$ interactions may occur.
Examples of these include the relatively nearby and better known
astrophysical sites such as our galaxy and the earth atmosphere,
where the basic $pp$ interactions occur as $pA$ interactions. The
$pp$ interactions in these sites form a rather certain background
to the extra-galactic high-energy tau neutrino searches. It is
possible that such interactions are the only sources of
high-energy tau neutrinos should the search for high-energy tau
neutrinos originating from several proposed distant sites such as
AGNs, GRBs, as well as groups and clusters of galaxies, turns out
to be negative. Therefore, it is rather essential to investigate
the high-energy tau neutrino flux expected from our galaxy and the
earth atmosphere.

Both the galactic and atmospheric tau neutrinos can be categorized
into intrinsic and oscillated ones. Here, intrinsic $\nu_{\tau}$
flux refers to the $\nu_{\tau}$ produced directly by an
interaction while oscillated $\nu_{\tau}$ refers to the
$\nu_{\tau}$ resulted from the $\nu_{\mu} \to \nu_{\tau}$
oscillation. Presently, there exists no estimate for the intrinsic
high-energy $\nu_{\tau}$ flux originating from our galaxy in $pp$
interactions, although the estimates for $\nu_e$ and $\nu_{\mu}$
fluxes due to the same interactions are available
\cite{Stecker:1978ah}. In this work, we calculate the intrinsic
$\nu_{\tau}$ flux from our galaxy by using the perturbative and
nonperturbative QCD approaches to model the $pp$ interactions, and
taking into account all major tau neutrino production channels up
to $E\leq 10^{11}$ GeV. We note that the production of tau
neutrinos in the terrestrial context was discussed in
Ref.~\cite{DeRujula:1992sn}, which uses a nonperturbative QCD
approach for $pp$ interactions. To calculate the oscillated
galactic and atmospheric $\nu_{\tau}$ fluxes, we apply the
two-flavor neutrino oscillation analysis \cite{Fukuda:2000np}.

It is essential to compare the above galactic tau neutrino flux
with the flux of atmospheric tau neutrinos. The intrinsic
atmospheric tau neutrino flux has been calculated for $E\leq 10^6$
GeV \cite{Pasquali:1998xf}. In this work, we extend the
calculation up to $E\leq 10^{11}$ GeV. Such an extension requires
the input of cosmic-ray flux spectrum for an energy range beyond
that considered in Ref.~\cite{Pasquali:1998xf}. Furthermore, for a
greater neutrino energy, the solutions of cascade equations
relevant to the neutrino production behave differently. For the
oscillated $\nu_{\tau}$ flux, it is interesting to note that the
oscillation length for $\nu_{\mu}\to \nu_{\tau}$ for the energy
range $10^{3} \leq E/\mbox{GeV} \leq 10^{11}$ is much greater than
the thickness of the earth atmosphere. Hence the oscillated
atmospheric $\nu_{\tau}$ flux in this case is highly suppressed.

One may argue that the interaction of the high-energy cosmic-rays
with the ubiquitous cosmic microwave background (CMB) photons
present in cosmos ($p\gamma $ rather than $pp$ interactions) could
also be an important source for high-energy astrophysical tau
neutrinos. The center-of-mass energy ($\sqrt{s}$) needed to
produce a $\tau$ lepton and a $\nu_\tau $ is  at least $\sim $ 1.8
GeV. In a collision between a proton with an energy $E_p$ and a
CMB photon with an energy $E_{\gamma_{\rm CMB}}$, the
center-of-mass energy squared of the system satisfies $m_p^2 < s <
4 E_p E_{\gamma_{\rm CMB}} + m_p^2$. Since the peak of the CMB
photon flux spectrum with a temperature $\sim $ 2.7 K is at about
$2.3\cdot 10^{-4}$ eV,  it requires a very energetic proton with
$E_p \agt 2.5\cdot 10^{12}$ GeV in order to produce a $\tau
\nu_\tau$ pair. Thus, the contribution of the intrinsic tau
neutrino flux from the interaction between the cosmic proton and
the CMB photon is negligible, unless we are considering extremely
high-energy protons, beyond the presently observed highest energy
cosmic-rays \cite{Nagano:ve}. To compute the oscillated
$\nu_{\tau}$ flux in this case, we consider the non-tau neutrino
flux generated by the $p\gamma$ interaction via the $\Delta $
resonance, commonly referred to as the Greisen-Zatsepin-Kuzmin
(GZK) cutoff interaction, which assumes that the proton travels a
cosmological distance \cite{Greisen:1966jv}. A recent calculation
of the intrinsic non-tau GZK neutrino flux indicates that this
flux peaks typically at $E\sim 10^{9}$ GeV \cite{Engel:2001hd},
beyond the reach of presently operating high-energy neutrino
telescopes such as AMANDA and Baikal experiments
\cite{Halzen:2001ty}. This flux decreases for $E< 10^{9}$ GeV. In
fact, it falls below the intrinsic non-tau galactic-plane neutrino
flux for $E\leq 5\cdot 10^{7}$ GeV. The neutrino flavor
oscillations of the intrinsic non-tau GZK neutrinos into tau
neutrinos result in a $\nu_{\tau}$ flux comparable to the original
non-tau neutrino flux \cite{Athar:2000je}. Therefore, in the
absence (or smallness) of tau neutrino flux from other possible
extra-galactic astrophysical sites, the only source of high-energy
tau neutrinos besides the atmospheric background is from our
(plane of) galaxy, typically for $10^{3} \leq E/\mbox{GeV} \leq
5\cdot 10^{7}$. This is an energy range to be explored by the
above high-energy neutrino telescopes in the near future.

The organization of the paper is as follows. In Section II, we
discuss the calculation of intrinsic high-energy tau neutrino flux
from our galaxy, including the description of the flux formula,
the galaxy model, and the various tau neutrino production channels
taken into account in the calculation. Although this flux will be
shown small, we shall go through some details of the calculation
since they are also useful for the calculation in the next
section. In Section III, we present our result on the intrinsic
atmospheric $\nu_{\tau}$ flux and compare it with the galactic
one. In Section IV, we discuss the effects of neutrino flavor
mixing, which are used to construct the oscillated $\nu_{\tau}$
fluxes from the galaxy and the earth atmosphere respectively. The
total galactic $\nu_{\tau}$ flux (the sum of intrinsic and
oscillated fluxes) is compared to its atmospheric counterpart, and
the dominant energy range for the former flux is identified. We
also mention the currently envisaged prospects for identifying the
high-energy tau neutrinos. We summarize in Section V.

\section{The intrinsic galactic tau neutrino flux}

\subsection{ The tau-neutrino flux formula and the galaxy model}

We use the following formula for computing the $\nu_{\tau}$ flux:
%
%
\begin{equation}
\label{AA}
 \frac{\mbox{d}N_{{\nu}_{\tau}}}{\mbox{d}E}=
 \int_{E}^{\infty}
 \mbox{d}E_p \; \phi_{p}(E_{p}) \, f(E_{p}) \;
 \frac{\mbox{d}n_{pp \to \nu_{\tau}+Y}}   {\mbox{d}E}.
\end{equation}
In the above equation, $E$ is the tau neutrino energy and the
cosmic-ray flux spectrum, $\phi_{p}(E_{p})$, is given by
\cite{JACEE}
%
%
\begin{equation}
\label{proton}
 \phi_{p}(E_{p})= \left \{
           \begin{array}{ll}
          1.7 \, (E_{p}/\mbox{GeV})^{-2.7} & {\rm for}\; E_{p}<E_{0}, \\
          174 \, (E_{p}/\mbox{GeV})^{-3}   & {\rm for}\; E_{p}\geq
          E_{0},
        \end{array}
      \right .
\end{equation}
where $E_{0}=5\cdot 10^{6}$ GeV and $\phi_{p}(E_{p})$ is in units
of cm$^{-2}$ s$^{-1}$ sr$^{-1}$ GeV$^{-1}$. We assume directional
isotropy in $\phi_{p}(E_{p})$ for the above energy range. The
function $f(E_{p})$ is equal to $R/\lambda_{pp}(E_{p})$,  where
$\lambda_{pp}(E_{p})=(\sigma^{{\rm incl}}_{pp}n_{p})^{-1}$ is the
$pp$ interaction length and $R$ is a representative distance in
the galaxy along the galactic plane. The  target particles are
taken to be protons with a constant number density of 1 cm$^{-3}$
and $R$ is taken to be $\sim 10$ kpc, where 1 pc $\simeq$ $3 \cdot
10^{18}$ cm. The $\sigma^{{\rm incl}}_{pp}$ is the total inelastic
$pp$ cross section. Since the high-energy protons traverse a
distance $R$ much shorter than the proton interaction length, the
proton flux spectrum $\phi_p(E_p)$ is assumed to be constant over
the distance $R$. Furthermore, we calculate the intrinsic tau
neutrino flux along the galactic plane only to obtain the maximal
expected tau neutrino flux. The matter density decreases
exponentially in the direction orthogonal to the galactic plane,
therefore the amount of intrinsic tau neutrino flux decreases by
approximately two orders of magnitude for the energy range of our
interest. For further details, see Ingelman and Thunman in Ref.
\cite{Stecker:1978ah}. The function $f(E_{p})$ in Eq. (\ref{AA})
basically gives the number of inelastic $pp$ collisions in the
distance $R$, while the distribution $\mbox{d} n/\mbox{d} E$ is
the differential cross section normalized by $\sigma^{{\rm
incl}}_{pp}$, i.e.,
%
%
\begin{equation}
\label{dnpp}
 \frac{\mbox{d}n_{pp \to \nu_{\tau}+Y}}   {\mbox{d}E}=
 \frac{1}{\sigma^{{\rm incl}}_{pp}}
 \frac{\mbox{d}\sigma_{pp \to \nu_{\tau}+Y}}   {\mbox{d}E} \;.
\end{equation}
The above distribution gives the fraction of inelastic $pp$
interactions that goes into $\nu_{\tau}$'s.  We can now simplify
Eq. (\ref{AA}) into
%
%
\begin{equation}
\label{dNtau}
 \frac{\mbox{d}N_{{\nu}_{\tau}}}{\mbox{d}E}=
 R n_p \int_{E}^{\infty}
 \mbox{d}E_p \; \phi_{p}(E_{p}) \,
 \frac{\mbox{d}\sigma_{pp \to \nu_{\tau}+Y}}   {\mbox{d}E} \;.
\end{equation}
It is clear that the task of computing
$\mbox{d}N_{{\nu}_{\tau}}/\mbox{d}E$  relies on the evaluation of
the differential cross section $\mbox{d}\sigma/ \mbox{d}E$ in $pp$
interactions. In this work, we include all major production
channels of tau neutrinos, namely, via the $D_s$ meson,
$b$-hadron, $t\bar t$, $W^*$, and $Z^*$. In general, the symbol
$\nu_{\tau}$ shall be used to account for the contribution of
$\nu_{\tau}$ and $\bar{\nu}_{\tau}$ unless otherwise mentioned. We
note that the heavy intermediate states, such as the  $D_{s}$
meson, $b$-hadrons and other heavier states, decay (into
$\nu_{\tau}$) before they interact with other particles in the
galactic plane. This is due to the rather small matter density of
the medium and the large distance between the proton source and
the earth. Before we proceed, let us remark that the intrinsic tau
neutrino production by the galactic $p\gamma $ interactions is
suppressed relative to that in the galactic $pp$ interaction,
because $n_{\gamma}\ll n_{p}$ for the energy range of interest to
us \cite{info}.

\subsection{Tau neutrino production}
\subsubsection{Via $D_s$ mesons}

The lightest meson that can decay into a $\tau$-$\nu_\tau$ pair is
the $D_s$ meson.  It was pointed out that the production and the
decay of $D_s$ meson is the major production channel for tau
neutrinos in the AGN \cite{Athar:1998ux}. We expect the same
will be true for galactic tau neutrinos.  The $D_s$ meson decays
into a charged $\tau$ lepton and a
 $\nu_\tau$.  The charged $\tau$ lepton subsequently decays into the second
$\nu_\tau$ plus other particles, which can be one prong or three
prong. The kinematics of the $\tau$ lepton decay is treated by the
Monte Carlo technique. For simplicity, we assume that the $\tau$
lepton  decays into a $\nu_\tau$ and a particle $Y$ with the mass
$m_Y$ satisfying $0.1\, {\rm GeV} < m_Y < m_\tau -0.1\,{\rm GeV}$.
Consequently, the second $\nu_\tau$ is much more $energetic$ than
the first one because the $D_s$ mass is only slightly larger than
$m_\tau$. We take the branching ratio $B(D_s \to \tau^+ \nu_\tau)$
as $\simeq 0.074$ \cite{Groom:in}.

We now turn to the production of $D_s$ mesons.  Here, we employ
two approaches to calculate the production of $D_s$ mesons: (i)
the perturbative QCD (PQCD) and (ii) the quark-gluon string model
(QGSM).  In the PQCD approach, we use the leading-order result for
$pp\to c \bar c$:
%
%
\begin{equation}
\label{pqcd}
 \sigma( pp \to c\bar c) = \sum_{ij} \, \int
\int \mbox{d} x_1 \mbox{d} x_2\; f_{i/p} (x_1) f_{j/p} (x_2) \,
\hat \sigma \left( ij \to c\bar c \right) \;,
\end{equation}
where $f_{i/p}(x)$ are the parton distribution functions (we use
the CTEQv5 \cite{Lai:1999wy}), while the parton subprocesses are
$q\bar q,gg \to c\bar c$.  We use a $K$ factor, $K=2$, to account
for the NLO corrections \cite{Nason:1989zy}. The matrix elements
for these subprocesses can be found in Appendix A.  The $c$ or
$\bar c$ then undergoes fragmentation into the $D_s$ meson, which
we model by the Peterson fragmentation function
\cite{Peterson:1982ak} with $\epsilon\approx 0.029$
\cite{Barate:1999bg} (see Appendix A). The probability $f_{c\to
D_s}$ of a $charm$ quark  fragmenting into a $D_s$ is 0.19
\cite{Barate:1999bg} (we have added the $f_{c\to D_s}$ and
$f_{c\to D_s^*}$). Typically, a fraction $z\;(z<1)$  of the
charm-quark energy is transferred to the $D_s$ meson so that the
energy spectrum of $D_s$ is softer than that of the charm quark.

The QGSM approach is nonperturbative and is based on the string
fragmentation. It contains a number of parameters determined by
experiments  \cite{Kaidalov:xg}. The production cross section of
the $D_s$ meson is given by the sum of $n$-pomeron terms
%
%
\begin{equation}
\label{qgsm} \frac{\mbox{d}\sigma^{D_s} (s,x)}{\mbox{d} x} \approx
\frac{1}{x^2 + x_\perp^2} \sum_{n=1}^\infty \; \sigma^{pp}_n (s)
\, \phi^{D_s}_n (s,x) \;,
\end{equation}
where $x=2 p_\parallel/\sqrt{s}$ and $x_\perp = 2\sqrt{ (m^2_{D_s}
+ p^2_\perp)/s}$. The functions $\sigma^{pp}_n(s)$ and
$\phi^{D_s}_n (s,x)$ are given in Appendix B.

\begin{figure}[ht!]
\includegraphics[width=5in]{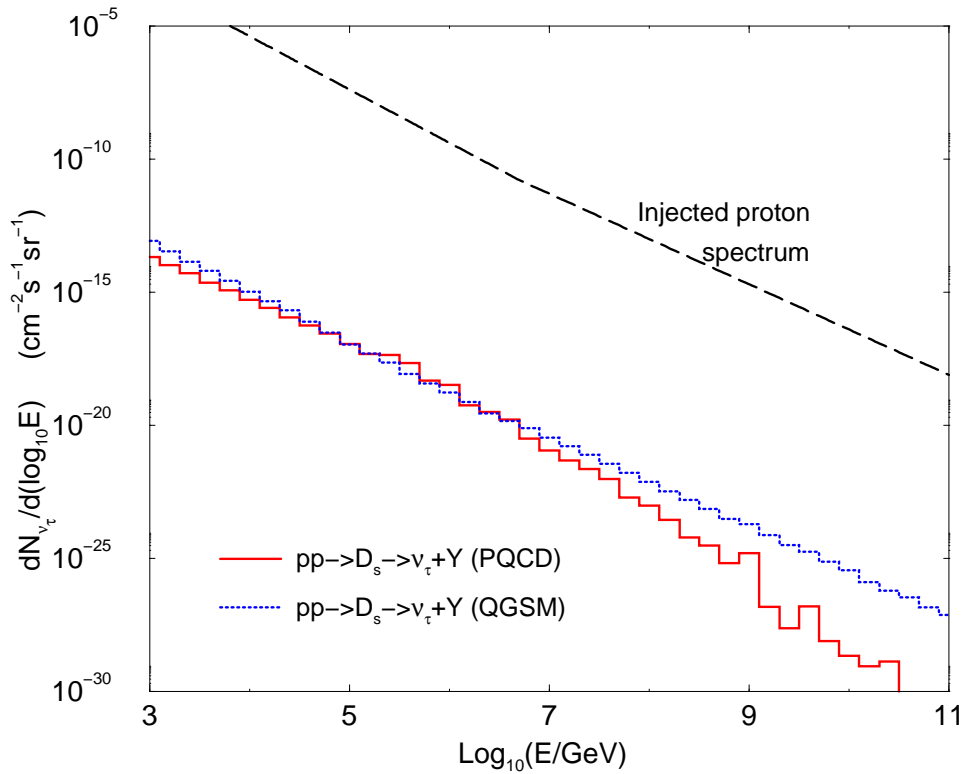}
\caption{\small \label{fig1} A comparison between the PQCD and
QGSM approach to the energy spectrum of the intrinsic galactic
$\nu_{\tau}$ flux coming from the $D_s$ meson. The thick dashed
curve is the injected proton flux spectrum given by Eq.
(\ref{proton}).}
\end{figure}

A comparison of these two approaches for $D_{s}$ meson is shown in
Fig. \ref{fig1}. The dashed line in the figure is the spectrum of
the injected proton flux given by  Eq. (\ref{proton}). The
$\nu_\tau$ spectra calculated by these two approaches agree well
with each other for $E\leq 10^{6}$ GeV.  Beyond this energy, the
QGSM approach gives a relatively harder spectrum.  In fact, this
behavior was already seen in the $\mbox{d}\sigma/\mbox{d}x$
distribution.
%
%
Nevertheless, in the
region where the two approaches differ, the tau neutrino flux is
already  small.
To our knowledge,  the heaviest meson production that the QGSM has
been applied to is the $D_s$ meson production.  The current
highest energy collider experiment
for $D_{s}$ meson production is at the FERMILAB TEVATRON
with $\sqrt{s}=1.8 \cdot 10^{3}$ GeV,
which corresponds to an $E_{p}\sim 1.7\cdot 10^6$~GeV, as $s\sim
2 m_{p}E_{p}$ in our setting. Note that up to this $\sqrt{s}$,
the agreement between the two approaches is quite good,
according to Fig. \ref{fig1}.
We have used a factorization scale $Q^2 =
\hat s /4$ and the one-loop running strong coupling constant $\alpha_s$
with the value $\alpha_s(Q^2=M_Z^2)=0.118$, and the $m_c = 1.35$~GeV.

An important quantity in the neutrino flux calculation is the
average fraction of the injected proton energy being transferred
to the tau neutrino, i.e., the ratio $r \equiv E/E_p$. The average
value of $r$ is given either by the mean
%
%
\begin{equation}
\label{averagey}
 \langle r \rangle = \frac{ \displaystyle{\int \,r
 \, \left(\frac{\mbox{d} \sigma}{\mbox{d} E} \right ) \, \mbox{d}r }}{
 \displaystyle{\int \,
 \left(\frac{\mbox{d}\sigma}{\mbox{d} E} \right ) \,\mbox{d}r
 }} \;,
\end{equation}
or by the value of $r$ at which the distribution $\mbox{d}
\sigma/\mbox{d} E$ attains the peak.  We have found that both
averages of $r$ are very close to each other. The $\langle r
\rangle $ ranges from $5 \cdot 10^{-3}$ to $5\cdot 10^{-7}$ for
$E_p$ from $10^{3}$ to $10^{11}$ GeV for the production channel
$pp\to c\bar c \to D_s  +Y \to \nu_\tau +Y$, using the PQCD
approach.  The higher the injected proton energy, the smaller is
the fraction of the incident $E_{p}$ that goes into hadrons.

\subsubsection{Via $b\bar{b}$, $t\bar{t}$, $W^*$ and $Z^*$}

The production of $b \bar b$ and $t \bar{t}$ in $pp$ interactions
can be calculated quite reliably by the PQCD approach, similar to
the calculation of $c\bar c$. The relevant matrix elements,
including the ones for $W^{*}$ and $Z^{*}$, are listed in Appendix
A.
%
\begin{figure}[ht!]
\includegraphics[width=5in]{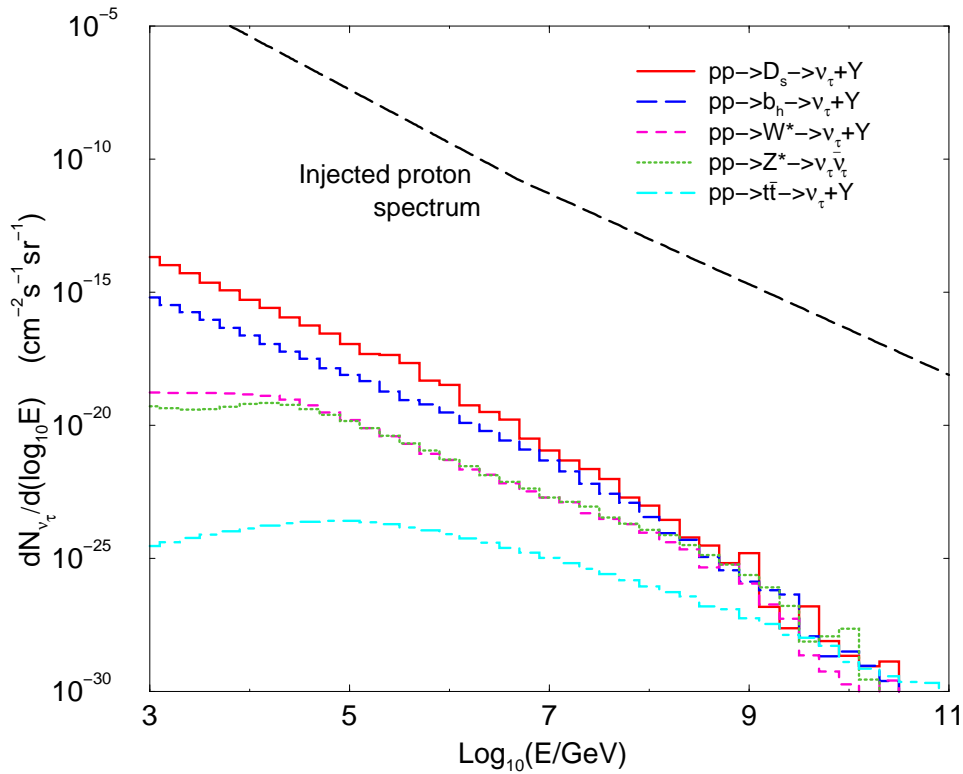}
\caption{\small \label{fig2} Intrinsic galactic tau neutrino flux
calculated via various intermediate states and channels: $D_s$,
$b$-hadron, $W^*$, $Z^*$, and $t\bar t$. The injected proton flux
spectrum is also shown. }
\end{figure}
The results \footnote{After taking into account the sources of
uncertainties such as values of $m_c$, $m_b$, $m_t$, $R$, $n_p$,
etc., we estimate that our calculation of the intrinsic galactic
$\nu_{\tau}$ flux is reliable within an order of magnitude.}are
shown in Fig. \ref{fig2}.  A few observations can be drawn from
the figure \cite{Cheung}. (i) The production via $D_s$ mesons
dominates for $E \leq 10^{9}$ GeV, followed by $b$-hadrons, $W^*$,
$Z^*$, and $t\bar t$ respectively. (ii) For $E\geq 10^{9}$ GeV all
these production channels become $comparable$. (iii) The intrinsic
tau neutrino flux is about $10-12$ orders of magnitude smaller
than the injected proton flux.

\section{The intrinsic atmospheric tau neutrino flux}

The earth atmosphere is an interesting extra-terrestrial site
where the basic $pp$ interaction occurs in the form of a $pA$
collision with $A$ the nuclei present in the earth atmosphere.
Incidently, it is the only known nearby extra-terrestrial site
from where the intrinsic neutrinos are observed as a result of
high-energy cosmic-ray interactions.

We have calculated the downward and horizontal intrinsic
atmospheric $\nu_{\tau}$ flux for the energy range $10^{3} \leq
E/\mbox{GeV} \leq 10^{11}$. We used the nonperturbative QCD
approach mentioned in the last section to model the production of
$D_{s}$ mesons in $pA$ interactions. We have used the
$\phi_{p}(E_{p})$ given by Eq. (\ref{proton}) and the $Z$-moment
description for the calculation of intrinsic tau neutrino flux,
which is appropriate for a varying target density medium
\cite{Gaisser:vg}. Since the tau neutrino flux is determined by
the flux of $D_s$ meson, we briefly discuss the cascade equation
for the $D_s$ flux. In general, we have \cite{Gaisser:vg}
\begin{equation}
\label{cascade}
\frac{\mbox{d}\phi_{D_s}}{\mbox{d}X}=-
\frac{\phi_{D_s}}{\lambda_{D_s}}
-\frac{\phi_{D_s}}{\rho d_{D_s}}+Z_{D_s D_s}\frac{\phi_{D_s}}{\lambda_{D_s}}
+Z_{p D_s}\frac{\phi_p}{\lambda_p},
\end{equation}
where $X$ is the slant depth, i.e., the amount of atmosphere (in
${\rm g}/{\rm cm}^2$) traversed by the $D_s$ meson ($X=0$ at the
top of the atmosphere), $\phi_p(E,X)$ and $\phi_{D_s}(E,X)$ are
the fluxes of protons and $D_s$ mesons respectively. The
$\lambda_{D_s}$ and $d_{D_s}$ are the interaction thickness (in
${\rm g}/{\rm cm}^2$) and the decay length of the $D_s$ meson
respectively. Finally, the $Z$-moments $Z_{p D_s}$ and $Z_{D_s
D_s}$ describe the effectiveness of generating $D_s$ meson from
the higher-energy protons and $D_s$ mesons respectively. We have
\begin{eqnarray}
Z_{pD_s}(E)&=&\int_{E}^{\infty} \mbox{d}E' \frac{\phi_p(E',0)}{\phi_p(E,0)}
\frac{\lambda_p(E)}{\lambda_p(E')}
\frac{\mbox{d}n_{pA\to D_s + Y}(E,E')}{\mbox{d}E},\nonumber \\
Z_{D_sD_s}(E)&=&\int_{E}^{\infty} \mbox{d}E' \frac{\phi_{D_s}(E',0)}{\phi_{D_s}(E,0)}
\frac{\lambda_{D_s}(E)}{\lambda_{D_s}(E')}
\frac{\mbox{d}n_{D_s A\to D_s + Y}(E,E')}{\mbox{d}E},
\end{eqnarray}
where $\mbox{d}n_{pA\to D_s +Y}/\mbox{d}E$ and $\mbox{d}n_{D_s
A\to D_s +Y}/\mbox{d}E$ are defined according to Eq.~(\ref{dnpp}).
We note that Eq.~(\ref{cascade}) should be solved together with
the cascade equation governing the propagation of high-energy
cosmic-ray protons. In fact, the proton flux equation can be
easily solved such that
\begin{equation}
\phi_p(E,X)\approx \exp(-\frac{X}{\Lambda_p})\,\phi_p(E,0),
\end{equation}
where $\Lambda_p\equiv \lambda_p/(1-Z_{pp})$ is the proton
attenuation length with $\lambda_p$ the proton interaction
thickness (in ${\rm g}/{\rm cm}^2$) and the $Z_{pp}$ given by
\begin{equation}
Z_{p p}(E)=\int_{E}^{\infty} \mbox{d}E'
\frac{\phi_p(E',0)}{\phi_p(E,0)}
\frac{\lambda_p(E)}{\lambda_p(E')}\frac{\mbox{d}n_{pA\to p+
Y}(E,E')} {\mbox{d}E}.
\end{equation}
An analytic solution of Eq.~(\ref{cascade}) can be obtained for
either the low or the high energy limit. Such limits are
characterized by whether the $D_{s}$ decays before it interacts
with the medium or vice versa. The critical energy separating the
two limits is approximately $\sim 8.5 \cdot 10^{7}$ GeV. In the
low energy limit, we disregard the first and third terms in the
R.H.S. of Eq.~(\ref{cascade}). On the other hand, one can drop the
second term in the high energy limit. With $\phi_{D_s}$
determined, the $\nu_{\tau}$ flux can be calculated by considering
the decay $D_s\to \nu_{\tau}\tau$ and the subsequent decay
$\tau\to \nu_{\tau} +Y$ as treated in Section II. We first
obtain two $\nu_{\tau}$ fluxes,
valid for low and high energy limits respectively, in terms of
$Z$-moments and then interpolate the two fluxes. A complete
numerical solution without the interpolation is given in a
separate work \cite{phd}.
%
%
\vspace{1.5cm}
\begin{figure}[ht!]
\includegraphics[width=5in]{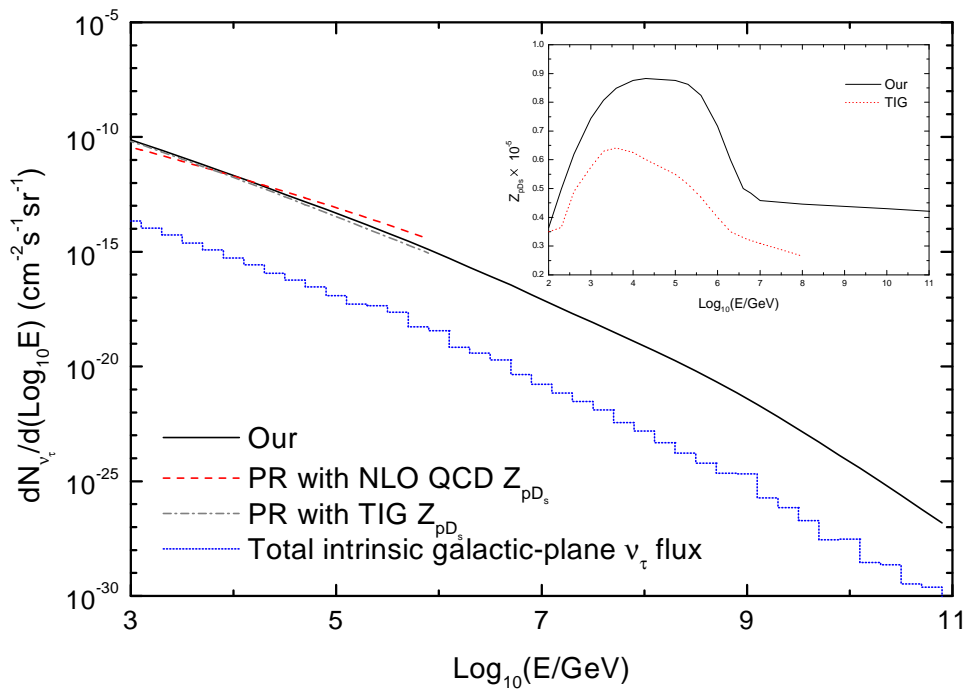}
\caption{\small \label{fig2'} Intrinsic horizontal $\nu_{\tau}$
flux via production and decay of the $D_s$ meson in the earth
atmosphere. For $10^{3} \leq E/\mbox{GeV} \leq 10^{6}$, the
results by PR are also shown. In the inset, we compare our
calculated $Z_{pD_{s}}$ with the one given by rescaling TIG's
$Z_{pD^0}$. The total intrinsic galactic-plane tau neutrino flux
is also shown.}
\end{figure}

In Fig. \ref{fig2'}, we show our result for the intrinsic
atmospheric $\nu_{\tau}$ flux along the horizontal direction. For
comparison, the results by Pasquali and Reno (PR)
\cite{Pasquali:1998xf}, valid for $10^{3} \leq E/\mbox{GeV} \leq
10^{6}$, are also shown. The $\nu_{\tau}$ flux along all the other
direction is small. For example\footnote{The upward going
high-energy $\nu_{\tau}$ with a energy $E\geq 10^4$ GeV is
degraded in energy to about $E\leq 10^3$ GeV after crossing the
earth.}, the downward $\nu_{\tau}$ flux is about $8$ times smaller
than the horizontal one for $E\geq 10^8$ GeV. We remark that the
major uncertainty for determining the above $\nu_{\tau}$ flux is
the $Z$-moment $Z_{pD_s}$. In Ref. \cite{Pasquali:1998xf}, the
authors calculate $Z_{pD_s}$ using two different approaches, which
then give rise to different results for the $\nu_{\tau}$ flux. The
first approach is based upon next-to-leading order (NLO)
perturbative QCD \cite{Frixone}, while the second approach
rescales the $Z_{pD^0}$ given by the PYTHIA \cite{PYTHIA}
calculation of Thunman, Ingelman, and Gondolo (TIG)
\cite{Gondolo:1995fq}. In the inset of Fig. \ref{fig2'} , we also
show our calculated $Z_{pD_{s}}$ in comparison with the one given
by rescaling TGI's result for $Z_{pD^0}$. We do not show the
$Z_{pD_{s}}$ obtained by NLO perturbative QCD since it is not
explicitly given in Ref.~\cite{Pasquali:1998xf}.

\section{Effects of oscillations and Prospects for observations}

 In the context of two neutrino flavors,
$\nu_{\mu}$ and $\nu_{\tau}$, the total $\nu_{\tau}$ flux,
 $\mbox{d}N^{\rm{tot}}_{\nu_{\tau}}/\mbox{d}(\log_{10}E)$, is
 given by \cite{Athar:2000yw}
%
%
\begin{equation}
\label{osc}
 \mbox{d}N^{\rm tot}_{\nu_{\tau}}/\mbox{d}(\log_{10}E)= P\cdot
 \mbox{d}N_{\nu_{\mu}}/\mbox{d}(\log_{10}E)+(1-P)\cdot
 \mbox{d}N_{\nu_{\tau}}/\mbox{d}(\log_{10}E).
\end{equation}
 Here $P\equiv P(\nu_{\mu}\to \nu_{\tau})=
 \sin^{2}2\theta \cdot \sin^{2} (l/l_{\rm osc})$. The neutrino flavor
oscillation length for $\nu_{\mu} \to \nu_{\tau}$ is $l_{{\rm
osc}}\sim (E/\delta m^{2})$. For $10^{3}\leq E/\mbox{GeV} \leq
10^{11}$ and with $\delta m^{2}\sim 10^{-3}$ eV$^{2}$, we obtain
$10^{-8}\leq l_{{\rm osc}}/\mbox{pc} \leq 1$. We assume maximal
flavor mixing between $\nu_{\mu}$ and $\nu_{\tau}$.

For intrinsic neutrinos produced along the galactic plane, we take
$\mbox{d}N_{\nu_{\mu}}/\mbox{d}(\log_{10}E)$ given by Ingelman and
Thunman in Ref. \cite{Stecker:1978ah} by extrapolating it up to $E
\leq 10^{11}$ GeV, whereas for
$\mbox{d}N_{\nu_{\tau}}/\mbox{d}(\log_{10}E)$, we use our results
obtained in Section II. For galactic-plane neutrinos, we note that
$l_{{\rm osc}}\ll l$, where $l\sim 5$ kpc is the typical average
distance the intrinsic high-energy muon neutrinos traverse after
being produced in our galaxy. Eq. (\ref{osc}) then implies that,
on the average, half of the muon neutrino flux will be oscillated
into tau neutrino flux, reducing its intrinsic level to one half.

\begin{figure}[ht!]
\includegraphics[width=5in]{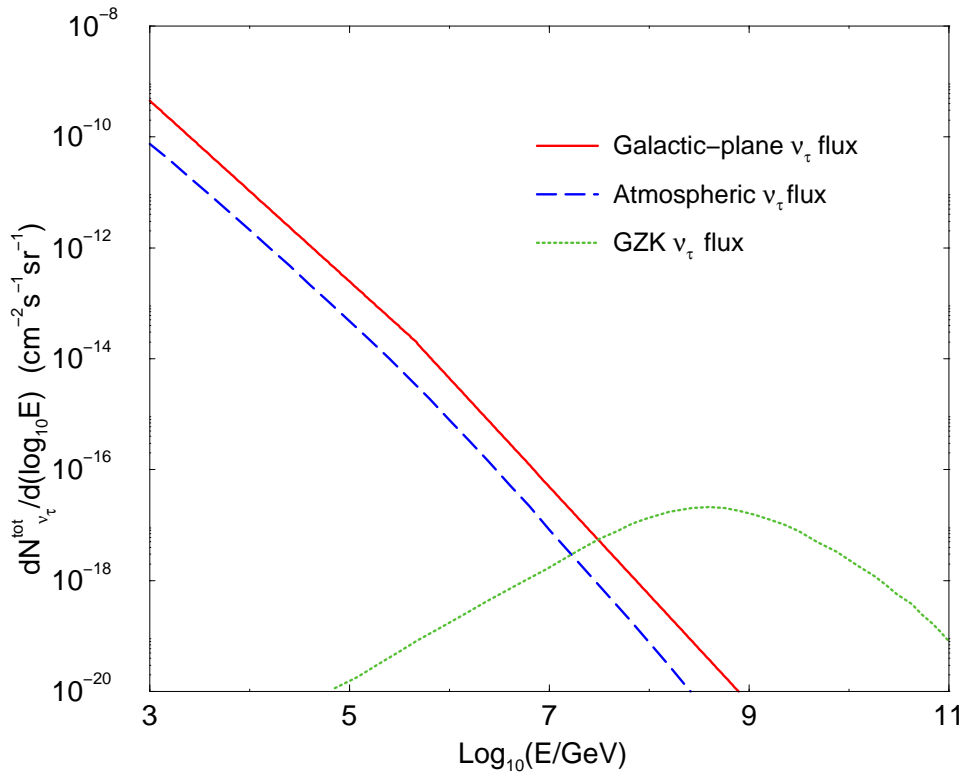}
\caption{\small \label{fig2''} Galactic-plane, horizontal
atmospheric and GZK tau neutrino fluxes under the assumption of
neutrino flavor oscillations.}
\end{figure}

For downward going neutrinos produced in the earth atmosphere, we
take $l \simeq 20$ km as an example. We use
 the (prompt) $\mbox{d}N_{\nu_{\mu}}/\mbox{d}(\log_{10}E)$ given in
\cite{Volkova:th}, whereas for
$\mbox{d}N_{\nu_{\tau}}/\mbox{d}(\log_{10}E)$, we use our results
obtained in Section III. For horizontal and upward going
atmospheric neutrinos, $l\sim 10^{3}-10^{4}$ km. Here, $l_{\rm
osc}\gg l$, and so the intrinsic atmospheric tau neutrino flux
dominates over the oscillated one for $E\geq 10^{3}$ GeV,
essentially irrespective of the incident direction. We present
these results in Fig. \ref{fig2''}, along with the GZK oscillated
tau neutrino flux briefly mentioned in Section I. For the GZK
neutrinos, $l\geq $ Mpc and
$\mbox{d}N_{\nu_{\mu}}/\mbox{d}(\log_{10}E)$ is taken from Ref.
\cite{Engel:2001hd}. From the figure, we note that the
galactic-plane oscillated $\nu_{\tau}$ flux {\em dominates} over
the intrinsic atmospheric $\nu_{\tau}$ flux for $E\leq 5\cdot
10^{7}$ GeV, whereas the GZK oscillated tau neutrino flux
dominates for $E\geq 5\cdot 10^{7}$ GeV.

A prospective search for high-energy tau neutrinos can be done by
appropriately utilizing the characteristic $\tau$ lepton range in
deep inelastic (charged current) tau-neutrino-nucleon
interactions, in addition to the attempt of observing the associated
showers. For $E$ close to $6\cdot 10^{6}$ GeV, the (anti electron)
neutrino-electron resonant scattering is also available to
the search for (secondary) high-energy tau neutrinos
\cite{Fargion:1997eg}. The main advantages of using the latter
channel are that the neutrino flavor in the initial state is least
affected by neutrino flavor oscillations and that this cross
section is free from theoretical uncertainties
\cite{Athar:2001bi}. The appropriate utilization of the
characteristic $\tau$ lepton range in both interaction channels  not
only helps to identify the incident neutrino flavor  but also
helps to bracket the incident neutrino energy as well, at least in
principle.

%
%

For downward going or near horizontal high-energy  tau neutrinos,
the deep inelastic neutrino-nucleon scattering, occurring near or
inside the detector, produces two (hadronic) showers
\cite{Learned:1994wg}. The first shower is due to a
charged-current neutrino-nucleon deep inelastic scattering,
whereas the second shower is due to the (hadronic) decay of the
associated $\tau$ lepton produced in the first shower. It might be
possible for the proposed large neutrino telescopes such as
IceCube to constrain the two showers simultaneously for
$10^{6}\leq E/\mbox{GeV} \leq 10^{7}$, depending on the achievable
shower separation capabilities \cite{Athar:2000ak} (see, also,
\cite{Halzen:2001ty}). Here, the two showers develop mainly in
ice. Using the same shower separation criteria as given in
\cite{Athar:2000ak}, we note that the $\sim $ (100 m)$^{3}$
proposed neutrino detector, commonly called the megaton detector
\cite{Chen:2001eq}, may constrain the two showers separated by
$\geq $ 10 m,  typically for $5\cdot 10^{5}\leq E/\mbox{GeV} \leq
10^{6}$. The two nearly horizontal showers may also possibly be
contained in a large surface area detector array such as  Pierre
Auger, typically for $5\cdot 10^{8}\leq E/\mbox{GeV}\leq 10^{9}$
\cite{Athar:2000tg}. In contrast to previous situations, here the
two showers develop mainly in air. Several different suggestions
have recently been made to measure only one shower, which is due
to the $\tau$ lepton decay, typically for $10^{8}< E/\mbox{GeV} <
10^{10}$, while the first shower is considered to be mainly
absorbed in the earth \cite{Bertou:2001vm}. The upward going
high-energy tau neutrinos with $E \geq 10^{4}$ GeV, on the other
hand, may avoid earth shadowing to a certain extent because of the
characteristic $\tau$ lepton range, unlike the upward going
electron and muon neutrinos, and may appear as a rather small pile
up of $\nu_{l}$ ($l=e, \, \mu, \, \mbox{and}\, \tau $) with $E
\sim 10^{3}$ GeV \cite{Halzen:1998be,Becattini:2000fj}. However,
the empirical determination of incident tau neutrino energy seems
rather challenging here.

The above studies indicate that for a rather large range of tau
neutrino energy, a prospective search may be carried out. The
event rate in each experimental configuration is directly
proportional to the incident $\nu_{\tau}$ flux and the effective
area of the detector concerned. Presently, no direct empirical
upper bounds (or observations) for high-energy tau neutrinos
exist.

To have an idea of the event rate, let us consider the downward
going high-energy tau neutrinos originating from the
galactic-plane due to neutrino flavor oscillations. The flux of
such neutrinos has been given in Fig. 4. Following Ref.
\cite{Athar:2000ak}, we note that these galactic tau neutrinos
give a representative event rate of  $\leq 1$  per year per
steradian for two separable and contained showers with $E\sim
10^{6}$ GeV in a km$^{3}$ volume neutrino telescope such as the
proposed IceCube.

\section{Discussion and Conclusions}

We have calculated the $\nu_{\tau}$ flux due to $pp$ interactions
in our galaxy.  This flux consists of intrinsic tau neutrino flux
and that arising from the oscillations of muon neutrinos. We note
that the latter flux is dominant over the former by four to five
orders of magnitude for the considered neutrino energy range. From
Fig. 4, one can see that the main background for the search of
high-energy extra-galactic tau neutrino is due to the muon
neutrinos produced in the galactic-plane, which then oscillate
into tau neutrinos.  Such a flux dominates for $10^{3} \leq
E/\mbox{GeV} \leq 5\cdot 10^{7}$. Therefore it is clear that
searching for extra-galactic tau neutrinos orthogonal to the
galactic-plane is more prospective.

In the calculation of galactic tau neutrino flux, we have used a
simplified model of matter distribution along our galactic plane
to obtain the maximal intrinsic tau neutrino flux. We have
explicitly calculated the contribution of heavier states such as
$b\bar{b}$, $t\bar{t}$ as well as $W^{*}$ and $Z^{*}$  in addition
to the more conventional $D_{s}$ channel to the intrinsic tau
neutrino flux. We have  estimated the average fraction of the
incident cosmic-ray energy that goes into tau neutrinos and found
it to be less than 1$\%$. The contributions from $b\bar{b}$,
$t\bar{t}$, $W^{*}$ and $Z^{*}$ channels are comparable to $D_s$
for $E \geq 10^{9}$ GeV. For $D_{s}$ channel, we have used both
perturbative and nonperturbative QCD approaches.

We have extended a previous calculation of intrinsic atmospheric
$\nu_{\tau}$ flux from $E \leq 10^{6}$ GeV up to $E \leq 10^{11}$
GeV. Here, we used the nonperturbative QCD approach to calculate
the production of $D_{s}$ mesons in $pA$ interactions. In
comparison with the intrinsic galactic-plane $\nu_{\tau}$ flux, it
is large. However, since the distance between the detector and the
neutrino source in the galactic plane is sufficiently large, the
neutrino flavor oscillations of non-tau neutrinos into tau
neutrinos makes the eventual tau neutrino flux along the galactic
plane greater than the atmospheric tau neutrino flux for $10^{3}
\leq E/\mbox{GeV} \leq 5\cdot 10^{7}$. However, the intrinsic
atmospheric $\nu_{\tau}$ flux dominates over the oscillated
galactic $\nu_{\tau}$ flux in the direction orthogonal to the
galactic plane. We have also briefly mentioned the presently
envisaged prospects for observations. In summary, we have
completed the compilation of all definite sources of tau neutrino
flux, i.e., those from our galaxy and from the earth atmosphere.
Such a compilation is needed before one conducts the search for
tau neutrinos from extra-galactic sources.

\acknowledgments H.A. and K.C. are supported in part by the
Physics Division of National Center for Theoretical Sciences under
a grant from the National Science Council of Taiwan. G.L.L. and
J.J.T. are supported by the National Science Council of R.O.C.
under the grant number NSC90-2112-M009-023.

\appendix

\section{Formulas for PQCD}

In this appendix, we list the matrix element squared for the
subprocesses of $pp\to Q\bar Q$, where $Q=c,b,t$, used in the PQCD
calculation.
\begin{eqnarray}
   \frac{\mbox{d} \hat \sigma}{\mbox{d} \cos\theta^*} (q\bar q\to Q \bar Q)&=&
   \frac{g_s^4 \beta}{72 \pi \hat s^{3}} \biggr[ (m_Q^2 -\hat t)^2 + (m_Q^2
   -\hat u)^2 + 2 \hat s m_Q^2 \biggr] \;, \nonumber\\
 \frac{\mbox{d} \hat \sigma}{\mbox{d} \cos\theta^*} (gg\to Q \bar Q)&=&
    \frac{g_s^4 \beta}{768 \pi \hat s} \biggr \{
    \frac{4}{(\hat t- m_Q^2)^2} ( -m_Q^4 -3m_Q^2 \hat t -m_Q^2\hat u +\hat
    u \hat t ) \nonumber \\
    &+& \frac{4}{(\hat u- m_Q^2)^2} ( -m_Q^4
    -3m_Q^2 \hat u -m_Q^2 \hat t + \hat u \hat t )  \nonumber \\
 &+& \frac{m_Q^2}{(\hat u-m_Q^2)(\hat t -m_Q^2)}( 2m_Q^2 + \hat t + \hat u )
    +18 \frac{1}{\hat s^2} ( m_Q^4 - m_Q^2 (\hat t+ \hat u) + \hat t \hat u )
         \nonumber \\
        &+& \frac{9}{\hat t -m_Q^2} \frac{1}{\hat s} ( m_Q^4 - 2 m_Q^2 \hat t +
        \hat u \hat t ) + \frac{9}{\hat u -m_Q^2} \frac{1}{\hat
        s} ( m_Q^4 - 2 m_Q^2 \hat u + \hat u \hat t ) \Biggr \},
\end{eqnarray}
where $g_s^2 = 4\pi \alpha_s$, $\beta = \sqrt{1-(4m^{2}_{Q}/\hat
s)}$ and $\hat s, \hat t, \hat u$ are the usual Mandelstem
variables with $\hat t-m_Q^2=-(\hat s/2) (1-\beta
\cos\theta^{*})$.

The subprocesses for high-energy tau neutrino production via
$W^{*}$ and $Z^{*}$  are $q\bar q' \to W^* \to \tau^\pm \nu_\tau$
and $q\bar q \to Z^* \to \nu_\tau \bar \nu_\tau$.  The spin- and
color-averaged amplitude squared for these subprocesses are given
by
\begin{eqnarray}
\overline{\sum} \left|{\cal M}(u \bar d \to W^* \to \tau^+
\nu_\tau) \right |
 &=& \frac{g^4}{12} \;  \frac{1}{ (\hat s - m_W^2)^2 + \Gamma_W^2 m_W^2 }\;
\hat t ( \hat t - m^2_\tau ), \nonumber \\
\overline{\sum} \left|{\cal M}(d \bar u \to W^* \to \tau^- \bar
\nu_\tau) \right |  &=& \frac{g^4}{12} \; \frac{1}{ (\hat s -
m_W^2)^2 + \Gamma_W^2 m_W^2 }\;
\hat u ( \hat u - m^2_\tau ), \nonumber \\
\overline{\sum} \left|{\cal M}(q \bar q \to Z^* \to \nu_\tau \bar
\nu_\tau) \right | &=& \frac{g^4}{3 \cos^4 \theta_{\rm w}}\;
\frac{1}{ (\hat s - m_Z^2)^2 + \Gamma_Z^2 m_Z^2 }\; \left[ \left(
g_L^\nu g_L^q \right)^2 \hat u^2 + \left ( g_L^\nu g_R^q\right)^2
\hat t^2 \right ]  \;,
\end{eqnarray}
where $g_L^f = T_{3f} - Q_f \sin^2\theta_{\rm w}$ and $\theta_{\rm
w}$ is the weak mixing angle, $T_{3f}$ is the third component of
the weak isospin and $Q_f$ is the electric charge in units of
proton charge of the fermion $f$.

The Peterson fragmentation function is given by
\begin{equation}
 D_{Q \to h_{Q}}(z)= N \frac{z(1-z)^{2}}{[(1-z)^{2}+\epsilon z]^{2}},
\end{equation}
where  $N$ is the normalization constant and $\epsilon $ is given
in the text.

\section{Formulas for QGSM}

 In Eq. (\ref{qgsm}), the functions $\sigma^{pp}_n(s)$ and
$\phi^{D_s}_n (s,x)$ are given as follows:

\begin{equation}
  \phi_n^{D_s}(s,x) = a_0^{D_s} \left [
    F_{qq}^{D_s}(x_+,n)F_{q}^{D_s}(x_-,n)
+ F_{q}^{D_s}(x_+,n)F_{qq}^{D_s}(x_-,n) + 2(n-1)
F_{sea}^{D_s}(x_+,n)F_{sea}^{D_s}(x_-,n) \right ],
\end{equation}
where
\begin{eqnarray}
F_{q}^{D_s}(x,n) &=& \frac{2}{3} \int_x^1 \mbox{d}x_1 \;
f_p^{u}(x_1,n) \;G_u^{D_s}
     \left(\frac{x}{x_1}\right)
+ \frac{1}{3} \int_x^1 \mbox{d}x_1 \;f_p^{d}(x_1,n) \; G_d^{D_s}
     \left(\frac{x}{x_1} \right), \\
F_{qq}^{D_s}(x,n) &=& \frac{2}{3} \int_x^1 \mbox{d}x_1
\;f_p^{ud}(x_1,n) \;G_{ud}^{D_s} \left(
       \frac{x}{x_1} \right )
+ \frac{1}{3} \int_x^1 \mbox{d}x_1 \;f_p^{uu}(x_1,n)
\;G_{uu}^{D_s} \left(
     \frac{x}{x_1} \right ), \\
 F_{sea}^{D_s}(x,n) &=& \frac{1}{4 + 2\delta_s + 2\delta_c} \Biggr \{
 \int_x^1 \mbox{d}x_1 \;f_p^{u_{sea}}(x_1,n) \;
    \left[ G_{u}^{D_s} \left( \frac{x}{x_1}\right) + G_{\bar{u}}^{D_s} \left(
          \frac{x}{x_1}\right ) \right] \nonumber\\
&&+ \int_x^1 \mbox{d}x_1 \;f_p^{d_{sea}}(x_1,n) \;
   \left[ G_{d}^{D_s} \left(\frac{x}{x_1}\right) + G_{\bar{d}}^{D_s} \left(
        \frac{x}{x_1} \right) \right] \nonumber \\
&&+ \delta_s \int_x^1 \mbox{d}x_1 \;f_p^{s_{sea}}(x_1,n)
 \;\left[ G_{s}^{D_s} \left(\frac{x}{x_1}\right) + G_{\bar{s}}^{D_s} \left(
       \frac{x}{x_1}\right ) \right] \nonumber\\
&&+ \delta_c \int_x^1 \mbox{d}x_1 \;f_p^{c_{sea}}(x_1,n) \;
 \left[ G_{c}^{D_s} \left(\frac{x}{x_1}\right) + G_{\bar{c}}^{D_s} \left(
     \frac{x}{x_1}\right) \right]  \Biggr \}.
\end{eqnarray}

In the above, $f_p^{i}(x,n)$'s are the distribution functions
describing the $n-$Pomeron distribution functions of quarks or
diquarks ($i=u,d,uu...$) with a fraction of energy $x$ from the
proton, and $G_{i}^{h}(z)$'s are the fragmentation functions of
the quark or diquark chain into a hadron $h$ which carries a
fraction $z$ of its energy.

The list of the $f_p^{i}(x,n)$ are given by
\begin{eqnarray}
  f_p^{u}(x,n) &=&
    \frac{\Gamma(1+n-2\alpha_N)}{\Gamma(1-\alpha_R) ~\Gamma(\alpha_R - 2\alpha_N + n)}
    \times x^{-\alpha_R} ~(1-x)^{\alpha_R - 2\alpha_N + (n-1)}, \nonumber\\
    &=& f_p^{u_{sea}}, \nonumber\\
  f_p^{d}(x,n) &=&
    \frac{\Gamma(2+n-2\alpha_N)}{\Gamma(1-\alpha_R) ~\Gamma(\alpha_R - 2\alpha_N + n+1)}
    \times x^{-\alpha_R} ~(1-x)^{\alpha_R - 2\alpha_N + n}, \nonumber\\
    &=& f_p^{d_{sea}}, \nonumber\\
  f_p^{uu}(x,n) &=&
    \frac{\Gamma(2+n-2\alpha_N)}{\Gamma(-\alpha_R + n) ~\Gamma(\alpha_R - 2\alpha_N + 1)}
\times x^{\alpha_R - 2\alpha_N + 1} ~(1-x)^{-\alpha_R + (n-1)}, \nonumber\\
  f_p^{ud}(x,n) &=&
    \frac{\Gamma(1+n-2\alpha_N)}{\Gamma(-\alpha_R + n) ~\Gamma(\alpha_R - 2\alpha_N + 2)}
    \times x^{\alpha_R - 2\alpha_N} ~(1-x)^{-\alpha_R + (n-1)}, \nonumber\\
  f_p^{s_{sea}}(x,n) &=&
    \frac{\Gamma(1 + n + 2\alpha_R - 2\alpha_N -2 \alpha_{\phi})}
    {\Gamma(1-\alpha_{\phi}) ~\Gamma(2\alpha_R - 2\alpha_N + n -\alpha_{\phi})}
    \times  x^{-\alpha_{\phi}} ~(1-x)^{2 \alpha_R - 2\alpha_N + (n-1) -\alpha_{\phi}}, \nonumber \\
  f_p^{c_{sea}}(x,n) &=&
    \frac{\Gamma(1 + n + 2\alpha_R - 2\alpha_N -2 \alpha_{\psi})}
        {\Gamma(1-\alpha_{\psi}) ~\Gamma(2\alpha_R - 2\alpha_N + n -\alpha_{\psi})}
    \times x^{-\alpha_{\psi}} ~(1-x)^{2 \alpha_R - 2\alpha_N + (n-1) -\alpha_{\psi}}, \nonumber
\end{eqnarray}
where $\Gamma $ is the usual Gamma function.

The list of the $G_{i}^{D_s^{\pm}}(z)$  are given by
\begin{eqnarray}
  G_{u,\bar{u},d,\bar{d}}^{D_s^{\pm}} (z)
&=& (1-z)^{\lambda-\alpha_{\psi} + 2 -\alpha_R - \alpha_{\phi}},  \nonumber\\
  G_{uu,ud}^{D_s^{\pm}} (z)
&=& (1-z)^{\lambda-\alpha_{\psi} +\alpha_R -2\alpha_N
-\alpha_{\phi} +2},
 \nonumber\\
  G_{s}^{D_s^{+}} (z)
    &=& (1-z)^{\lambda-\alpha_{\psi} + 2(1- \alpha_{\phi})}, \nonumber\\
    &=& G_{\bar{s}}^{D_s^{-}}(z), \nonumber\\
  G_{s}^{D_s^{-}} (z)
    &=& (1-z)^{\lambda-\alpha_{\psi}} \times (1 + a_1 ~z^2),  \nonumber\\
    &=& G_{\bar{s}}^{D_s^{+}}(z), \nonumber\\
  G_{c,\bar{c}}^{D_s^{\pm}} (z)
    &=& z^{1-\alpha_{\psi}} (1-z)^{\lambda-\alpha_{\phi}}. \nonumber
\end{eqnarray}

In the above, the input parameters are as follows:
\begin{eqnarray}
  \alpha_R &=& 0.5 ~,\nonumber\\
  \alpha_N &=& -0.5 ~,\nonumber\\
  \alpha_{\phi} &=& 0 ~,\nonumber\\
  \alpha_{\psi} &=& -2.18 ~,\nonumber\\
  \lambda &=& 0.5 ~,\nonumber\\
  \delta_s &=& 0.25 ~,\nonumber\\
  a_0^{D_s} &=& 0.0007 ~,\nonumber\\
  a_1 &=& 5 ~,\nonumber \\
  \delta_c &=& 0 \; (0.01), \;\;\; \mbox{if charm sea contribution is turned
off (on).} \nonumber
\end{eqnarray}

The function $\sigma_n^{pp} (s)$ is given by the following
formulas:
\begin{equation}
 \sigma_n^{pp}(\xi) = \frac{\sigma_p}{n~z} \;\left
     ( 1-\exp(-z)\sum_{k=0}^{n-1} ~\frac{z^k}{k!} \right) \;,
\end{equation}
where
\begin{eqnarray}
\xi &=& \ln \left(\frac{s}{1 ~(\mbox{GeV})^2}\right), \nonumber\\
z &=& \frac{2 C ~\gamma_p}{R^2 + {\alpha_p}^{\prime} ~\xi}
~\exp(\xi \Delta),
 \nonumber\\
 \sigma_p &=& 8\pi \gamma_p ~\exp(\xi \Delta)\;. \nonumber
\end{eqnarray}
The best fit parameters are as follows :

(i) for $\sqrt{s} ~\leq~ 10^3$ GeV
\begin{eqnarray}
  \gamma_p          &=& 3.64 ~ ~(\mbox{GeV})^{-2}, \nonumber \\
  R^2               &=& 3.56 ~ ~(\mbox{GeV})^{-2}, \nonumber \\
  \alpha_p^{\prime} &=& 0.25 ~ ~(\mbox{GeV})^{-2}, \nonumber \\
  C                 &=& 1.5, \nonumber \\
  \Delta            &=& 0.07. \nonumber
\end{eqnarray}

\bigskip

(ii) for $\sqrt{s} ~\geq~ 10^3$ GeV

\begin{eqnarray}
  \gamma_p          &=& 1.77 ~ ~(\mbox{GeV})^{-2}, \nonumber \\
  R^2               &=& 3.18 ~ ~(\mbox{GeV})^{-2}, \nonumber \\
  \alpha_p^{\prime} &=& 0.25 ~ ~(\mbox{GeV})^{-2}, \nonumber \\
  C                 &=& 1.5, \nonumber \\
  \Delta            &=& 0.139. \nonumber
\end{eqnarray}


\begin{thebibliography}{99}
%
\bibitem{Halzen:2001ty}
For a recent review article, see, for instance, F.~Halzen,
arXiv:astro-ph/0111059 and references therein.
%
\bibitem{Fukuda:1998mi}
Y.~Fukuda {\it et al.}  [Super-Kamiokande Collaboration],
Phys.\ Rev.\ Lett.\  81 (1998) 1562.
%
\bibitem{Fukuda:2000np}
S.~Fukuda {\it et al.}  [Super-Kamiokande Collaboration],
Phys.\ Rev.\ Lett.\  85 (2000) 3999.
%
\bibitem{Kodama:2000mp}
K.~Kodama {\it et al.}  [DONUT Collaboration],
Phys.\ Lett.\ B 504 (2001) 218;
J.~Sielaff  [DONUT Collaboration],
arXiv:hep-ex/0105042.
%
\bibitem{Stecker:1978ah}
F.~W.~Stecker,
Astrophys.\ J.\  228 (1979) 919;
G.~Domokos, B.~Elliott and S.~Kovesi-Domokos,
J.\ Phys.\ G 19 (1993) 899;
V.~S.~Berezinsky, T.~K.~Gaisser, F.~Halzen and T.~Stanev,
Astropart.\ Phys.\ 1 (1993) 281;
G.~Ingelman and M.~Thunman,
arXiv:hep-ph/9604286.
%
\bibitem{DeRujula:1992sn}
A.~De Rujula, E.~Fernandez and J.~J.~Gomez-Cadenas,
Nucl.\ Phys.\ B 405 (1993) 80;
M.~C.~Gonzalez-Garcia and J.~J.~Gomez-Cadenas,
Phys.\ Rev.\ D 55 (1997) 1297.
%
\bibitem{Pasquali:1998xf}
L.~Pasquali and M.~H.~Reno,
Phys.\ Rev.\ D 59 (1999) 093003.
%
\bibitem{Nagano:ve}
M.~Nagano and A.~A.~Watson,
Rev.\ Mod.\ Phys.\ 72 (2000) 689.
%
\bibitem{Greisen:1966jv}
K.~Greisen,
Phys.\ Rev.\ Lett.\ 16 (1966) 748;
G.~T.~Zatsepin and V.~A.~Kuzmin,
JETP Lett.\ 4 (1966) 78 [Pisma Zh.\ Eksp.\ Teor.\ Fiz.\ 4 (1966)
114].
%
\bibitem{Engel:2001hd}
R.~Engel, D.~Seckel and T.~Stanev,
Phys.\ Rev.\ D 64 (2001) 093010.
%
\bibitem{Athar:2000je}
H.~Athar,
arXiv:hep-ph/0008121.
%
\bibitem{JACEE}
T. H. Burnett {\em et al.} [JACEE Collaboration], Astrophys. J.
Lett. 349 (1990) L25.
%
\bibitem{info}
For an estimate of galactic $n_{\gamma}$, see, Berezinsky {\em et
al.,} in Ref. \cite{Stecker:1978ah}.
%
\bibitem{Athar:1998ux}
H.~Athar,
Nucl.\ Phys.\ Proc.\ Suppl.\ 76 (1999) 419.
%
\bibitem{Groom:in}
D.~E.~Groom {\it et al.}  [Particle Data Group Collaboration],
Eur.\ Phys.\ J.\ C 15 (2000) 1.
%
\bibitem{Lai:1999wy}
H.~L.~Lai {\it et al.}  [CTEQ Collaboration],
Eur.\ Phys.\ J.\ C 12 (2000) 375.
%
\bibitem{Nason:1989zy}
See, for instance, P.~Nason, S.~Dawson and R.~K.~Ellis,
Nucl.\ Phys.\ B 327 (1989) 49 [Erratum-ibid.\ B 335 (1989) 260].
%
\bibitem{Peterson:1982ak}
C.~Peterson, D.~Schlatter, I.~Schmitt and P.~M.~Zerwas,
Phys.\ Rev.\ D 27 (1983) 105.
%
\bibitem{Barate:1999bg}
R.~Barate {\it et al.}  [ALEPH Collaboration],
Eur.\ Phys.\ J.\ C  16 (2000) 597.
%
\bibitem{Kaidalov:xg}
A.~B.~Kaidalov,
Phys.\ Lett.\ B 116 (1982) 459;
A.~B.~Kaidalov and O.~I.~Piskunova,
Sov.\ J.\ Nucl.\ Phys.\  43 (1986) 994 [Yad.\ Fiz.\  43 (1986)
1545];
G.~G.~Arakelian and P.~E.~Volkovitsky,
Z.\ Phys.\ A 353 (1995) 87;
G.~H.~Arakelian,
Phys.\ Atom.\ Nucl.\  61 (1998) 1570 [Yad.\ Fiz.\ 61 (1998) 1682];
G.~H.~Arakelian and S.~S.~Eremian,
Phys.\ Atom.\ Nucl.\  62 (199) 1724 [Yad.\ Fiz.\ 62 (1999) 1851].
 See, also,
M.~C.~Gonzalez-Garcia and J.~J.~Gomez-Cadenas,
 in Ref. \cite{DeRujula:1992sn}.
%
\bibitem{Cheung}
For a more detailed discussion, see K.~ Cheung, in: H.~Athar,
G.-L.~Lin and K.-W.~Ng (Eds.), Proceedings of the First NCTS
Workshop on Astroparticle Physics, 2001, Kenting (Taiwan) [to be
published].
%
\bibitem{Gaisser:vg}
T.~K.~Gaisser, Cosmic Rays And Particle Physics, Cambridge
University Press, NewYork, 1990.
%
\bibitem{phd}
J.-J.~Tseng, Ph.D. Thesis, NCTU, 2002 (in preparation).
%
\bibitem{Frixone}
S. Frixione, M.~L.~ Mangano, P.~Nason, and G.~Ridolfi, Nucl.\ Phys.\ B 431 (1994) 453.
%
\bibitem{PYTHIA}
T. Sjostrand, Comput.\ Phys.\ Commun.\ 82 (1994) 74.
%
\bibitem{Gondolo:1995fq}
M.~Thunman, G.~Ingelman and P.~Gondolo,
Astropart.\ Phys.\ 5 (1996) 309.
%
\bibitem{Athar:2000yw}
See, for instance, H.~Athar, M.~Jezabek and O.~Yasuda,
Phys.\ Rev.\ D 62 (2000) 103007 and references therein.
%
\bibitem{Volkova:th}
L.~V.~Volkova and G.~T.~Zatsepin,
Phys.\ Atom.\ Nucl.\  64 (2001) 266 [Yad.\ Fiz.\ 64 (2001) 313].
%
\bibitem{Fargion:1997eg}
D.~Fargion,
arXiv:astro-ph/9704205.
%
\bibitem{Athar:2001bi}
H.~Athar and G.-L.~Lin,
arXiv:hep-ph/0108204; {\em ibid},
arXiv:hep-ph/0201026.
%
\bibitem{Learned:1994wg}
J.~G.~Learned and S.~Pakvasa,
Astropart.\ Phys.\ 3 (1995)  267.
%
\bibitem{Athar:2000ak}
H.~Athar,
Astropart.\ Phys.\ 14 (2000) 217;
H.~Athar, G.~Parente and E.~Zas,
Phys.\ Rev.\ D 62 (2000) 093010.
%
\bibitem{Chen:2001eq}
See, for instance, H.~Chen {\it et al.}  [VLBL Study Group H2B-1
Collaboration],
arXiv:hep-ph/0104266.
%
\bibitem{Athar:2000tg}
H.~Athar,
arXiv:hep-ph/0004083.
%
\bibitem{Bertou:2001vm}
D.~Fargion,
arXiv:astro-ph/0002453;
X.~Bertou, P.~Billoir, O.~Deligny, C.~Lachaud and
A.~Letessier-Selvon,
arXiv:astro-ph/0104452;
J.~L.~Feng, P.~Fisher, F.~Wilczek and T.~M.~Yu,
arXiv:hep-ph/0105067;
A.~Kusenko and T.~Weiler,
arXiv:hep-ph/0106071.
%
\bibitem{Halzen:1998be}
F.~Halzen and D.~Saltzberg,
Phys.\ Rev.\ Lett.\ 81 (1998) 4305.
%
\bibitem{Becattini:2000fj}
F.~Becattini and S.~Bottai,
Astropart.\ Phys.\ 15 (2001) 323;
S.~Iyer Dutta, M.~H.~Reno and I.~Sarcevic,
arXiv:hep-ph/0110245 and references  therein;
J.~F.~Beacom, P.~Crotty and E.~W.~Kolb,
arXiv:astro-ph/0111482.
%
\end{thebibliography}
\end{document}